\documentstyle{article}

\newskip\humongous \humongous=0pt plus 1000pt minus 1000pt
\def\caja{\mathsurround=0pt}
\def\eqalign#1{\,\vcenter{\openup1\jot \caja
	\ialign{\strut \hfil$\displaystyle{##}$&$
	\displaystyle{{}##}$\hfil\crcr#1\crcr}}\,}
\newif\ifdtup

\def\oldreffmt#1{\rlap{[#1]} \hbox to 2\parindent{}}

\def\figfmt#1{\rlap{Figure {#1}} \hbox to 1in{}}

\def\beq{\begin{equation}}
\def\eeq{\end{equation}}
\def\eq{\beq\eeq}
\relax

\def\pol{\varepsilon}

\def\c{\,\cdot\,}

\def\L{\left(}\def\R{\right)}

\def\spa#1.#2{\left\langle#1\,#2\right\rangle}
\def\spb#1.#2{\left[#1\,#2\right]}
\def\lor#1.#2{\left(#1\,#2\right)}
\def\sand#1.#2.#3{%
\left\langle\smash{#1}{\vphantom1}^{-}\right|{#2}%
\left|\smash{#3}{\vphantom1}^{-}\right\rangle}
\def\sandp#1.#2.#3{%
\left\langle\smash{#1}{\vphantom1}^{-}\right|{#2}%
\left|\smash{#3}{\vphantom1}^{+}\right\rangle}
\def\sandpp#1.#2.#3{%
\left\langle\smash{#1}{\vphantom1}^{+}\right|{#2}%
\left|\smash{#3}{\vphantom1}^{+}\right\rangle}
\catcode`@=11  
\def\meqalign#1{\,\vcenter{\openup1\jot\m@th
   \ialign{\strut\hfil$\displaystyle{##}$ && $\displaystyle{{}##}$\hfil
             \crcr#1\crcr}}\,}
\catcode`@=12  

\def\L{\left(}
\def\R{\right)}

\def\c{\mskip 1mu\cdot\mskip 1mu }

\def\eps{\epsilon}

\def\pol{\varepsilon}

\def\dl^#1_#2{\delta^{#1}{}_{#2}}

\def\Gbd{\dot G_B}
\def\Gbdd{\ddot G_B}

\catcode`@=11  
\def\meqalign#1{\,\vcenter{\openup1\jot\m@th
   \ialign{\strut\hfil$\displaystyle{##}$ && $\displaystyle{{}##}$\hfil
             \crcr#1\crcr}}\,}
\catcode`@=12  

\font\tenbf=cmbx10
\font\tenrm=cmr10
\font\tenit=cmti10
\font\elevenbf=cmbx10 scaled\magstep 1
\font\elevenrm=cmr10 scaled\magstep 1
\font\elevenit=cmti10 scaled\magstep 1

\font\ninerm=cmr9

\font\sevenrm=cmr7

\newcount\eq
\def\adv{\global\advance\eq by1}
\def\set#1#2{\setbox#1=\hbox{#2}}

\newcount\eqncount

\def\equn{
\global\advance\eqncount by1
\eqno{(\the\eqncount)}            }
\def\put#1{\global\edef#1{(\the\eqncount)}           }

\global\newcount\refno \global\refno=1
\newwrite\rfile
\newlinechar=`\^^J
\def\ref#1#2{\nref#1{#2}}
\def\nref#1#2{\xdef#1{\the\refno}%
\ifnum\refno=1\immediate\openout\rfile=refs.tmp\fi%
\immediate\write\rfile{{\noexpand\item \noexpand#1 .\ }#2.}%
\global\advance\refno by1}

\def\e{\eps}
\def\Det{\hat\Delta}

\def\al#1{\alpha_{#1}}

\begin{document}
\renewcommand{\thefootnote}{\alph{footnote}}


\ref\Long
{Z. Bern and D. A. Kosower, Nucl.\ Phys.\ {\elevenbf B379} (1992) 451}

\ref\Pascos
{Z. Bern and D. A. Kosower, in {\elevenit Proceedings of the PASCOS-91
Symposium}, eds.\ P. Nath and S. Reucroft;
Z. Bern, Pitt-92-05, to appear in Phys. Lett. B}

\ref\Scherk{J. Scherk, Nucl.\ Phys.\ {\elevenbf B31} (1971) 222;
A. Neveu and J. Scherk, Nucl.\ Phys.\ {\elevenbf B36} (1972) 155;
M.\ B.\ Green, J.\ H.\ Schwarz and L.\ Brink, Nucl.\ Phys.\
{\elevenbf B198} (1982) 472}

\ref\PreviousPent
{G. 't Hooft and M. Veltman, Nucl. Phys. {\elevenbf B153} (1979) 365;
W. van Neerven and J. A. M. Vermaseren,
Phys. Lett. {\elevenbf 137B} (1984) 241}

\ref\Mapping
{Z. Bern, D. C. Dunbar, Nucl.\ Phys.\ {\elevenbf B379} (1992) 562}

\ref\Background {L.F.\ Abbott, Nucl.\ Phys.\ {\elevenbf B185} (1981) 189}

\ref\AGS
{L.F\ Abbott, M.T.\ Grisaru and R.K.\ Schaeffer,
Nucl.\ Phys.\ {\elevenbf B229} (1983) 372}

\ref\GN {J.L.\ Gervais and A. Neveu, Nucl.\ Phys.\
{\elevenbf B46} (1972) 381}

\ref\MPX
{M.\ Mangano, S. Parke, and Z.\ Xu, Nucl.\ Phys.\ {\elevenbf B298} (1988) 653;
M.\ Mangano and S.J.\ Parke, Phys.\ Rep.\ {\elevenbf 200} (1991) 301}

\ref\Color {Z. Bern and D.A.\ Kosower, Nucl.\ Phys.\
{\elevenbf B362} (1991) 389}

\ref\XZC{%
F.\ A.\ Berends, R.\ Kleiss, P.\ De Causmaecker, R.\ Gastmans, and T.\ T.\ Wu,
        Phys.\ Lett.\ {\elevenbf 103B} (1981) 124;
P.\ De Causmaeker, R.\ Gastmans,  W.\ Troost, and  T.\ T.\ Wu,
Nucl. Phys. {\elevenbf B206} (1982) 53;
R.\ Kleiss and W.\ J.\ Stirling,
   Nucl.\ Phys.\ {\elevenbf B262} (1985) 235;
   J.\ F.\ Gunion and Z.\ Kunszt, Phys.\ Lett.\ {\elevenbf 161B} (1985) 333;
Z.\ Xu, D.-H.\ Zhang and L. Chang, Nucl.\ Phys.\ {\elevenbf B291} (1987) 392}

\ref\Grav {F.A. Berends, W.T. Giele, H. Kuijf,
Phys. Lett. {\elevenbf 211B} (1988) 91}

\ref\Elsewhere
{Z. Bern, L. Dixon and D.A. Kosower, in preparation}

\ref\Susy{%
M.T. Grisaru, H.N. Pendleton, Nucl.\ Phys.\ {\elevenbf B124} (1977) 81;
S.J. Parke, T.R. Taylor,  Phys.\ Lett.\
{\elevenbf 157B} (1985) 81}


\rightline{SLAC--PUB--6012}
\rightline{UCLA/92/TEP/47}
\rightline{CERN-TH.6733/92}

\begin{center}{{\tenbf THE FIVE GLUON AMPLITUDE AND ONE-LOOP INTEGRALS}%
\footnote
{\ninerm Talk presented by Z.B. at the DPF92 Conference,
      Fermilab, Nov.\ 9-14, 1992 \hfil}

\vglue .7cm
{\tenrm ZVI BERN \\}
\baselineskip=11pt
{\tenit UCLA, Los Angeles, CA 90024\\}
\vglue 0.4cm
{\tenrm LANCE DIXON\\}
{\tenit SLAC, P.O. Box 4349, Stanford, CA 94309\\}
\vglue 0.2cm
{\tenrm and}
\vglue 0.2cm
{\tenrm DAVID A. KOSOWER\\}
{\tenit CERN, Geneva, Switzerland and CE Saclay,
F-91191 Gif-sur-Yvette cedex, France\\}
\vglue 0.6cm
{\tenrm ABSTRACT}}
\end{center}
{\rightskip=3pc
\leftskip=3pc
\tenrm\baselineskip=12pt
\noindent
We review the conventional field theory description of the string motivated
technique.  This technique is applied
to the one-loop five-gluon amplitude.  To
evaluate the amplitude a general method for computing
dimensionally regulated one-loop integrals is outlined including results
for one-loop integrals required for the pentagon diagram and beyond.  Finally,
two five-gluon helicity amplitudes are given.
\vglue 0.6cm}
{\elevenbf\noindent 1. Introduction}
\vglue 0.2cm

\elevenrm
\baselineskip=17 pt
The search for new physics at current and future hadron colliders
demands that we first refine our understanding of events
originating in known physics, most importantly QCD-associated
background processes.
Because the perturbation expansion for jet physics
in QCD is not an expansion strictly in the coupling constant, but is rather
an expansion
in the coupling constant times various infrared logarithms, loop
corrections play an important role in matching theoretical expectations
to experimental data.   Thus far, the one-loop corrections are known only
for the most basic processes, matrix elements with four external partons.

In order to minimize the algebra required for one-loop computations involving
{\large $n$} external gluons
string motivated rules were developed in ref.~[\Long].
Although the method was originally derived from string theory, it has
been summarized in terms of simple rules which require no knowledge of
string theory$^{\Long,\Pascos}$. Since
string theories contain gauge theories in the infinite string tension
limit$^{\Scherk}$ and have a simpler organization of the amplitudes
than field theories, a string motivated organization of
the amplitude is more compact than a traditional Feynman diagram
organization.

Here we discuss the application of the string motivated
technique to the computation of the five-gluon amplitude.
This requires the evaluation of dimensionally regularized
pentagon integrals.
The computation of pentagon integrals in the
case in which all internal lines are massive has been
discussed by various authors$^{\PreviousPent}$.
In particular
van~Neerven and Vermaseren
have provided an efficient method for calculating such integrals
in four dimensions.
The techniques of van~Neerven and Vermaseren do not apply directly to
dimensionally-regularized integrals, however, and the required
pentagon integrals have not yet been presented in a closed and
useful form, which is to say with all poles in {\large$\eps = (4-D)/2$}
 manifest,
and with all functions of the kinematic invariants expressed in terms of
(poly)logarithms.
Here we will provide a formula which yields
such expressions.\footnote{
   We have been informed that R.K. Ellis, W. Giele and E. Yehudai
   have recently
   evaluated the pentagon integrals by an independent technique.}
We also present a general solution for one-loop integrals
beyond the pentagon.

\vglue 0.5cm
{\elevenbf\noindent 2. Review of String Motivated Methods}
\vglue 0.3cm

The string motivated rules evaluate a one-loop {\large $n$} gluon amplitude
in terms of substitution rules acting on a basic kinematic
expression.  In refs.~[\Long,\Pascos] the substitution rules
necessary to obtain the values of all diagrams associated with a
one-loop {\large $n$}-gluon amplitude have already been given.  Here we will
not present the rules but will instead briefly review the interpretation
of the rules in terms of conventional field theory.  The conventional
field theory ideas necessary to reproduce the simplicity of the string
are$^{\Mapping}$:

\noindent {\elevenit a) Use of background field gauge.} Calculations in QCD
have traditionally been performed in ordinary Feynman gauge.  Perhaps
a reason why the background field method$^{\Background}$ has not
been used for gluon amplitudes is that it inherently
seems to be a method for computing effective actions and not
scattering amplitudes.  However, as
shown a number of years ago by Abbott, Grisaru and Schaeffer$^{\AGS}$,
the background field method can in fact be used for
{\large$S$}-matrix computations; one simply sews trees onto the loops in some
other gauge to obtain the {\large$S$}-matrix elements.  In the background
field Feynman gauge, vertices can be organized to mimic the simple
structure inherent in the string motivated rules leading to large
simplifications for QCD amplitudes.  A convenient gauge for sewing
trees onto the loops is the non-linear Gervais-Neveu gauge$^{\GN}$
(which was also motivated by string theory) since it has
simple vertices.

\noindent
{\elevenit b) Color ordering of vertices.} This
amounts to rewriting the Yang-Mills structure constant as
{\large$f^{abc} = - i {\rm Tr}([T^a, T^b] T^c)/\sqrt{2}$}
and then evaluating the coefficient of a single color trace ordering$^{\MPX}$.
String theory motivates the use of a {\large$U(N_c)$} gauge group instead of
an {\large$SU(N_c)$} gauge group; the extra {\large$U(1)$} decouples but the
relevant color algebra is much simpler for {\large$U(N_c)$}. A full
description of the one-loop color decomposition has been given in
ref.~[\Color].

\noindent {\elevenit c) Systematic organization of the vertex algebra.} In
order to minimize the work involved, it is important to organize the
vertex algebra in a particular systematic fashion.
Because of the way in which the loop momentum enters
into the background field vertices it turns out that the integration
over loop momentum is trivial.  In conventional gauges or other
background field gauges the loop momentum enters into the vertices in
a much more complicated way, not allowing a simple systematic
organization.
Once the amplitude has been written in a form where the loop momentum
is integrated out, one can use the spinor helicity method$^{\XZC}$ to simplify
the expressions.

Given this field theory understanding of the string motivated method
one might conclude that string theory is no longer required. This,
however, misses the point behind the use of string theory; the point
is that string theory guides computational organizations of gauge
theory amplitudes where efficient organizations of the amplitude are
unknown (as the one-loop case was prior to string motivated methods).
Further examples where string theory provides useful insight which
would be difficult to obtain by conventional means are extensions to
multi-loops and calculations of gravity amplitudes$^{\Grav}$.  Even
with all known field theory tricks such as spinor helicity methods and
special gauge choices, in a conventional framework it is difficult to
envision the compact organization of the amplitudes implied by string
theory.

\vglue 0.5cm
{\elevenbf\noindent 3. One-Loop Integrals.}
\vglue 0.3cm

\def\sq{\mskip-1mu}
The integral we wish to evaluate is the {\large$n$}-gon with general
kinematics, whose momentum-space definition is
$$
  I_n[\tilde P(p_\mu)] =  (-1)^{n+1} (4 \pi)^{2-\eps}i
  \int \!{d^{4 -2\eps} p \over (2 \pi)^{4-2\eps}}
  { \tilde P(p_\mu) \over
\bigl( p^2-M_1^2 \bigr) \bigl( (p\!-\!p_1)^2 -M_2^2 \bigr)
    \cdots \bigl( (p\!-\!p_{n-1})^2 - M_n^2 \bigr) } \; .
\equn\put\BasicIntegral
$$
where {\large$\tilde P(p_\mu)$} is
a polynomial in the loop momentum and where we take
{\large$p_i^\mu\ \equiv\ \sum_{j=1}^i k_j^\mu$} and
{\large$ k_i^2 = m_i^2$}
with the {\large$k_i$} momenta of external particles.
For QCD, {\large $m_i = M_i = 0$}.
In the five point case, after Feynman parametrization the
integral becomes
$$
\hskip -.3 cm
I_5[ P(\{a_i\})] =
\!\int_0^1 \!\! d^5a_i\ {  \Gamma(3+\eps) P(\{a_i\})
\delta (1-{\textstyle \sum_i} a_i)
\over \Bigl[- {{{5\atop\sum}\atop{i=1}}}
 \Bigl( (s_{i,i\!+\!1} \sq -\sq M_i^2\sq -\sq M_{i\!+\!2}^2) a_ia_{i\!+\!2}
 + (m_i^2\sq -\sq M_i^2\sq -\sq M_{i\!+\!1}^2) a_i a_{i\!+\!1}  - M_i^2 a_i^2
   \Bigr)\Bigr]^{3+\eps} } \ .
\equn
$$
Following 't~Hooft and Veltman$^{\PreviousPent}$,
we make the change of variables
{\large$a_i = \al{i} u_i/ \sum^n_{j=1} \al{j} u_j$}, (no sum on {\large$i$})
where
$$
s_{i,i+1} - M_i^2 - M_{i+2}^2 \ =\ -{1\over\alpha_i\alpha_{i+2}}\ ,\qquad
  m_i^2 - M_i^2 - M_{i+1}^2
  \ =\ -{\hat m_i^2\over\alpha_i\alpha_{i+1}}\ ,\qquad
   M_i^2\ =\ -{\hat M_i^2\over\alpha_i^2}\ ,
\equn\put\RescaledVariables
$$
so that
\vskip -5mm
$$
I_5[ P(\{a_i\}] = \Gamma(3+\e)
  \int_0^1 d^5u\; { P(\{\alpha_i u_i\})\,\delta\L \sum u_i - 1\R\,
              \L\sum^5_{j=1} \alpha_j u_j\R^{1-m+2\eps}
 \over \left[ \sum_{i=1}^5 \Bigl(u_iu_{i+2} + \hat m_i^2 u_iu_{i+1}
- \hat M_i^2 u_i^2 \Bigr) \right]^{3+\eps}}
 \ .
\equn
$$
The key observation is that this integral can be expressed in terms of
derivatives acting on the scalar integral
$$
I_5[ P_m(\{a_i\}] = {\Gamma(2-m+2\eps)\over\Gamma(2+2\eps)}
         \, :  P_m\left(\left\{ \alpha_i
  {\partial\over\partial \alpha_i} \right\}\right) : \, I_5[1]
\equn\put\DiffFormula
$$
where {\large$ P_m$} is a homogeneous polynomial of degree {\large$m$}
and the normal
ordering signifies that all the {\large$\alpha_i$} should be brought
to the left of the derivatives.
This equation and its extensions forms the basis for obtaining all the
tensor integrals as derivatives of the scalar integral
and for a differential equation method for evaluating
integrals$^{\Elsewhere}$.

In order to use this equation we need the solution of the pentagon
scalar integral.   The general recursive
solution for {\large$n\ge5$} external legs is
$$
  I_n[1]\ =\ {1\over 2N_n} \Biggl[ {1\over 2} \sum_{i=1}^n \alpha_i
{\partial \Det_n \over \partial \alpha_i}\Bigr|_{\hbox{\sevenrm non}-\alpha_i\
\hbox{\sevenrm variables\ fixed}} \times
   I_{n-1}^{(i)}[1]\ +\ (n-5+2\e) \,\Det_n \, I_n^{D=6-2\e}[1] \Biggr]
\equn\put\GeneralSolution
$$
where {\large$I_{n-1}^{(i)}[1]$} is the {\large$(n-1)$}-point scalar integral
obtained by removing the
internal propagator between legs {\large$i-1$} and {\large$i$}
(defined mod {\large$n$}),
{\large$\Det_n = \det (2 k_i \c k_j)\times \prod_{i=1}^n \alpha_i^2$} is
the rescaled Gram determinant (with {\large$i,j = 1\cdots, n-1$}).
For {\large$n=5$} the {\large$\alpha_i$}
are defined in Eq.~\RescaledVariables.  In this case
the last term in Eq.~\GeneralSolution\
is suppressed by a power of {\large$\eps$}
since the {\large$D=6-2\eps$}
scalar pentagon, {\large$I_5^{D=6-2\e}[1]$}, is completely finite.  Thus, the
explicit value of
{\large$I_5^{D=6-2\e}[1]$} is not needed.  (It
also turns out that it is not needed when applying the
differentiation formula \DiffFormula .)
For {\large$n >5$} the last term vanishes for four-dimensional
external kinematics
due to the vanishing of the Gram determinant.
In this way we obtain a recursive
solution for all one-loop scalar integrals in terms of the box
integrals (which are relatively easy to evaluate).
The overall normalization is
{\large$ N_n = 2^{n-1} \det \rho$}
where
{\large$ \rho_{ij} = - {\textstyle {1 \over 2}} ((p_{j-1} - p_{i-1})^2 -
M_i^2 - M_j^2 ) \alpha_i \alpha_j $}
is independent of the {\large$\alpha_i$} when converted to rescaled variables
analogous to the pentagon ones in Eq.~\RescaledVariables. For the
massless pentagon {\large$N_5 = 1$}.
This solution
extends van Neerven and Vermaseren's$^{\PreviousPent}$
four-dimensional pentagon solution
to dimensional regularization and arbitrary numbers
of external legs.

One way to verify the solution \GeneralSolution\ for a particular
{\large$n$} is by considering the integral \BasicIntegral\ with an
inverse propagator in the numerator.  This integral can be evaluated
as either an {\large$n$}-point integral or as an {\large$(n-1)$}-point
integral.  By comparing the two forms and summing over cyclic
permutations with coefficients obtained from the solution
\GeneralSolution\ one can verify its correctness after using
{\large$\sum_{i=1}^n a_i =1$}.  A general proof will be given
elsewhere$^{\Elsewhere}$.

\vglue 0.5cm
{\elevenbf\noindent 4. Explicit Results for Amplitudes}
\vglue 0.3cm

 For five-point amplitudes a straightforward use of the spinor
helicity method is cumbersome.  By rewriting ratios of spinor inner
products in terms of more conventional kinematic variables and the
square-root of the pentagon Gram determinant the spinor helicity
method becomes more usable$^{\Elsewhere}$.   The five gluon amplitudes
can then be obtained by applying the string
motivated techniques and using the solution for the pentagon integral.
By following this procedure we have obtained the two finite five-gluon
one-loop {\large$SU(N_c)$} partial amplitudes:
$$
\eqalign{
 A_{5;1}^{\rm 1-loop}&(1^+,2^+,3^+,4^+,5^+)
=\ \Bigl(1+ {1\over 2} N_s^{\rm adj} -
 N_{\! f}^{\rm adj} \Bigr) \cr
& \hskip 2 cm \times
{i\over 48\pi^2}\,
  {  \spa1.2\spb1.2\spa2.3\spb2.3 + \spa4.5\spb4.5\spa5.1\spb5.1
   + \spa2.3\spa4.5\spb2.5\spb3.4
 \over \spa1.2 \spa2.3 \spa3.4 \spa4.5 \spa5.1 } \cr
 A_{5;1}^{\rm 1-loop}&(1^-,2^+,3^+,4^+,5^+)
=\ \Bigl(1+ {1\over 2} N_s^{\rm adj} -
 N_{\! f}^{\rm adj} \Bigr) \cr
& \hskip 2 cm \times
{ i\over 48\pi^2}\,
{1\over\spa3.4^2 }
 \biggl[ { \spb2.5^3 \over \spb1.2\spb5.1 }
 + { \spa1.4^3\spb4.5\spa3.5 \over \spa1.2\spa2.3\spa4.5^2 }
 - { \spa1.3^3\spb3.2\spa4.2 \over \spa1.5\spa5.4\spa3.2^2 } \biggr]
  \cr }
\equn
$$
where {\large$N_s^{\rm adj}$} and {\large$N_{\! f}^{\rm adj}$}
are the number of adjoint massless real scalars and
Weyl fermions. (Fundamental representation
fermions or scalars require an additional factor of {\large$1/N_c$}.)
We follow the same notation and normalizations
as in refs.~[\Long,\Pascos].
The corresponding double trace {\large$A_{5;3}$} partial
amplitudes follow from the formulae in ref.~[\Color].
These amplitudes satisfy the relevant supersymmetry Ward
identities$^{\Susy}$
(which are satisfied trivially
 since they hold for the integrand of each string motivated diagram).
The remaining helicity amplitudes will be presented elsewhere.

\vglue .2 cm
In summary, the string motivated organization of the {\large$n$}-gluon
amplitude plays a major role in simplifying the computation of the
five-gluon one-loop amplitude.  Additional ingredients which allow the
computation to be performed are simple formulae for the relevant
one-loop integrals and a rewriting of spinor-helicity invariants in
terms of more conventional kinematic quantities.  These issues will be
presented in detail elsewhere$^{\Elsewhere}$.

\vglue .3 cm
The work of Z.B. was supported by the Texas National Research Laboratory
Commission grant FCFY9202 while the work of L.D. was supported by the
US Department of Energy under grant DE-AC03-76SF00515.

\vglue 0.5cm
{\elevenbf\noindent 5. References \hfil}

\baselineskip 13 pt
\begin{description}
\immediate\closeout\rfile
\input refs.tmp\vfill\eject
\end{description}

\end{document}